\documentclass[10pt, a4paper, conference]{IEEEtran}

\usepackage{amsmath,amssymb}
\usepackage[linesnumbered,ruled]{algorithm2e}
\usepackage{graphicx,xcolor,array}
\usepackage{pgfplots}
\pgfplotsset{compat=1.14}
\usepackage{tikz}
\usetikzlibrary{arrows}
\usetikzlibrary{automata,positioning}
\usepackage{makecell}
\usepackage{url}
\PassOptionsToPackage{bookmarks={false}}{hyperref}
\usepackage{hyperref}
\usepackage{pifont}%
\renewcommand\IEEEkeywordsname{Keywords}
\newcommand{\cmark}{\ding{51}}
\newcommand{\xmark}{\ding{55}}
\newtheorem{conjecture}{Conjecture}

\begin{document}
\title{Trends in Development of Databases and Blockchain}
	
\author{
\IEEEauthorblockN{
    Mayank Raikwar\IEEEauthorrefmark{1},
    Danilo Gligoroski\IEEEauthorrefmark{1},
    Goran Velinov\IEEEauthorrefmark{2}
    }\\
\vspace{-0.2cm}
\IEEEauthorblockA{\IEEEauthorrefmark{1} Norwegian University of Science and Technology (NTNU)
Trondheim, Norway\\}
\IEEEauthorblockA{\IEEEauthorrefmark{2} University Ss. Cyril and Methodius Skopje, Macedonia\\
Email: \{mayank.raikwar,danilog\}@ntnu.no, goran.velinov@finki.ukim.mk}}

\maketitle

\begin{abstract}
This work is about the mutual influence between two technologies: Databases and Blockchain. It addresses two questions: 1. How the database technology has influenced the development of blockchain technology?, and 2. How blockchain technology has influenced the introduction of new functionalities in some modern databases? For the first question, we explain how database technology contributes to blockchain technology by unlocking different features such as ACID (Atomicity, Consistency, Isolation, and Durability) transactional consistency, rich queries, real-time analytics, and low latency. We explain how the CAP (Consistency, Availability, Partition tolerance) theorem known for databases influenced the DCS (Decentralization, Consistency, Scalability) theorem for the blockchain systems. By using an analogous relaxation approach as it was used for the proof of the CAP theorem, we postulate a  "DCS-satisfiability conjecture." For the second question, we review different databases that are designed specifically for blockchain and provide most of the blockchain functionality like immutability, privacy, censorship resistance, along with database features.
\end{abstract}


\vspace{-0.2cm}
\section{Introduction}
\vspace{-0.1cm}
Blockchain has gained immense popularity from the last decade, acting as a distributed ledger for peer to peer transactions in a secure and immutable way. The development of blockchain pushed the market of decentralized applications in varied enterprises such as financial markets, insurance industries, supply chain industry. In this decentralized network of peers, each peer has a replica of data. This decentralization of blockchain raises many questions on ``How and where to store data". The problem where to store blockchain data has been somehow solved by using decentralized cloud storage solutions, but these solutions suffer from limited capability and user privacy matters.

After the invent of bitcoin~\cite{Nakamoto_bitcoin}, many questions about the blockchain have been carried out as ``blockchain as a database" or ``difference between blockchain and database,"~\cite{Chowdhury2018}. Blockchain differs from traditional databases in numerous ways like its decentralization, cryptographic security using chained hashes, no administration control, immutability, freedom to transfer without the permission of any central authority. To cherish these differences, many enterprise applications upgraded their traditional database storage solution with blockchain to make their implementation more secure, involving less trust among the parties of the industry. Despite having the features mentioned above, blockchain still lacks some features which traditional database has. Blockchain can leverage the traditional database features by either integrating the traditional database with blockchain or, to create a blockchain-oriented distributed database. The inclusion of the database features will leverage the blockchain with low latency, high throughput, fast scalability, and complex queries on blockchain data. 
Thus having the features of both blockchain and database, the application enhances its efficiency and security. Many of the blockchain platforms are now integrating with a database.

In recent years, many blockchain databases have been developed and introduced. These distributed databases have their consensus mechanism for the joint agreement on a data block by the network parties. These blockchain databases support features like complex data types, rich query structure, ACID compliant~\cite{brewer2000towards}, low latency, fast scalability, and cloud hosting. The adoption of database features in blockchain or vice-versa is an interesting research topic. Few industries have already built their blockchain database with all the required features. Many companies, including database giants IBM, Oracle, and SAP, as well as startups such as FlureeDB~\cite{FlureeDB}, BigchainDB~\cite{BigchainDB}, have devoted their efforts to develop blockchain database solutions to support SQL-like queries. 

\begin{table*}[!ht]
    \centering
    \begin{tabular}{|c|c|c|c|}
    \hline
    \parbox{5.5cm}{\centering Feature} & \parbox{4.0cm}{\centering Database domain} & \parbox{1.0cm}{\centering \vspace{1.0mm}Influence direction\vspace{1.0mm}} & \parbox{5.5cm}{\centering Blockchain domain}\\
    \hline
    \parbox{5.5cm}{\vspace{1.0mm} High throughput and scalability \vspace{1.0mm}} & \parbox{4.0cm}{$\checkmark$ (in distributed databases)} & $\rightarrow$  & \parbox{5.5cm}{\centering To be implemented } \\
    \hline
    \parbox{5.5cm}{\vspace{1.0mm}Transactions latency \vspace{1.0mm}} & \parbox{4.0cm}{\centering Low} & $\rightarrow$  & \parbox{5.5cm}{\centering High} \\
    \hline
    \parbox{5.5cm}{\vspace{1.0mm}Serializable isolation} & \parbox{4.0cm}{\vspace{1.0mm} Alternatives to 2-phase locking \vspace{1.0mm}} & $\leftarrow$  & $\checkmark$  \\
    \hline
    \parbox{5.5cm}{\vspace{1.0mm} ACID properties \vspace{1.0mm}} & $\checkmark$  & $\rightarrow$  & \parbox{5.5cm}{\vspace{1.0mm}Hyperledger Fabric~\cite{Androulaki2018} due to CouchDB~\cite{Anderson2010}} \\
    \hline
    \parbox{5.5cm}{\vspace{1.0mm}Complex queries on the historic data \vspace{1.0mm}} & $\checkmark$  & $\rightarrow$  & \parbox{5.5cm}{\vspace{1.0mm} Techniques such as VQL~\cite{VQL} } \\
    \hline
    \parbox{5.5cm}{Decentralization} & \parbox{4.0cm}{\vspace{1.0mm} New, blockchain-style databases \vspace{1.0mm}} & $\leftarrow$  & $\checkmark$  \\
    \hline
    \parbox{5.5cm}{\vspace{1.0mm} Immutability (tamper-resistance) \vspace{1.0mm}} & \parbox{4.0cm}{\vspace{1.0mm} Mechanisms that prevents deletes and record updates' history \vspace{1.0mm}} & $\leftarrow$  & $\checkmark$  \\
    \hline
    \parbox{5.5cm}{\vspace{1.0mm} Movement of digital assets \vspace{1.0mm}} & \parbox{4.0cm}{\vspace{1.0mm} New, blockchain-style distributed databases \vspace{1.0mm}} & $\leftarrow$  & $\checkmark$  \\
    \hline
    \parbox{5.5cm}{\vspace{1.0mm} \scriptsize{CAP (Consistency, Availability, Partition tolerance) \vspace{1.0mm}}} & $\checkmark$  & $\rightarrow$  & \parbox{5.5cm}{\vspace{1.0mm}\scriptsize{DCS (Decentralization, Consistency, Scalability) \vspace{1.0mm}}} \\
    \hline
    \end{tabular}\vspace{0.5mm}
    \caption{A summary of the mutual influence and the entangled development of Databases and Blockchain}
    \label{tab:InfluenceEntanglement}
    \vspace{-0.9cm}
\end{table*}

\textbf{Our Contribution.} In the last decade, we witnessed a tremendous interchanged and mutually influenced development of the database and blockchain technologies.  A performance study on distributed database~\cite{Bergman2018} for blockchain has already been done, but that involves very few databases. However, to the best of our knowledge, no work has been done towards a systematized study of those development trends. This work is neither about the specific cryptographic characteristics and components of the blockchain systems (that can be found in numerous surveys or systematization of knowledge studies~\cite{raikwar2019sok}) nor about the specific use of blockchain in some specific industries~\cite{hasselgren2019blockchain}. This work is about providing a detailed summary of traditional databases that are used or can be used in the design of blockchain platforms or applications. The work is also about a detailed explanation of different decentralized solutions that use traditional databases but provide blockchain-enabled solutions. Finally, this paper describes the DCS theorem and the trade-off properties present in the blockchain systems in a similar way to the CAP theorem for the database systems~\cite{gilbert2002brewer}. We hope that our work will be useful for industries or academia within the blockchain as a guide for choosing the appropriate database for their blockchain use-cases from the list of databases mentioned. Additionally, using our work, those involved in the research and the development of modern databases can potentially upgrade the functionalities of the databases that they are developing with some blockchain functionalities.

\vspace{-0.3cm}
\section{Mutual influence and development} \label{Motivation}
\vspace{-0.1cm}
 
Blockchain and database both can achieve many functionalities and features by coping with each other. If we frame blockchain as a database to provide a storage mechanism, then we can analyze how it differs from actual database systems. The following are the key points where blockchain and database differ in their properties, but both can leverage and enhance the characteristics of each other.
\begin{itemize}
    \item Traditional blockchain throughput decreases when the processing capacity of nodes participating in the blockchain increases. Yet, in the case of the distributed database, the throughput increases when the nodes increases. Hence throughput can be enhanced.
    \item The latency of transactions in blockchain is usually high compared to the latency in database. Thus, the latency can be made low as desired with the use of a database.
    \item Transactions in blockchain require serializable isolation, which can be achieved by consensus algorithms providing strong consistency. For the databases, there is a well-understood mechanism called 2-phase locking and concurrency-control~\cite{Kung1981}. However, new blockchain databases such as BlockchainDB~\cite{El-Hindi2019} based on MongoDB~\cite{MongoDB17} start to offer new transaction mechanisms based on blockchain.
    \item Most of the blockchain platforms do not support complex queries in its historic data. These queries are needed in many applications to retrieve the desired information. The complex query feature is available in most of the databases, but the provenance queries on historic data can be supported by the use of Multi-Version Concurrency Control~\cite{Bernstein1983}.
    \item The decentralization feature of blockchain has rewired most of the financial systems and industries from the last decade. Decentralization is not available in the traditional distributed database. With the advent of new blockchain-style databases, the decentralization is now possible and leads a promising growth to be used in many applications.
    \item One of the other excellent features of blockchain is immutability or tamper-resistance of transactions. This tamper-resistance can be achieved in database systems by mechanisms that disallow the deletes and updates in the database.
    \item Blockchain allows the creation and movement of digital assets, which is not allowed in a classical database. But, a blockchain-style distributed database can have this feature as a built-in feature.
\end{itemize}
\vspace{-0.15cm}

In Table \ref{tab:InfluenceEntanglement} we give a summary of this mutual influence and the entangled development of databases and blockchain.  

\vspace{-0.2cm}
\section{CAP theorem for blockchain} \label{CAP}
\vspace{-2.0mm}
CAP was introduced 20 years ago by Brewer~\cite{brewer2000towards} as a principle or conjecture, and two years later, it was proven in the asynchronous network model as a CAP theorem by Gilbert and Lynch in \cite{gilbert2002brewer}. In the same paper, similar impossibility results were proven for a partially synchronous network model. Additionally, by weakening the consistency conditions, they showed that it is possible to achieve all three properties in the so-called \textit{$t$-Connected Consistency} model. 

In more details, CAP theorem identifies the three specific system properties for any distributed/decentralized system. These properties are \textbf{C}onsistency, \textbf{A}vailability and \textbf{P}artition Tolerance.
\begin{itemize}
    \item \textit{Consistency} - Any read in the distributed system gives the latest write on the nodes.
    \item \textit{Availability} -  A Client always receives a response at any point of time irrespective of whether the read is the latest write.
    \item \textit{Partition Tolerance} - In case of partition between nodes in the distributed system, the system should still be functioning.
\end{itemize}

CAP theorem states that it is possible to achieve two of these three properties as guaranteed features in a distributed network, but it is impossible to achieve all three features at the same time. In practice, a distributed system always needs to be partition tolerant, thus leaving us to choose one property from Consistency or Availability. Hence, there is a trade-off between consistency and availability.

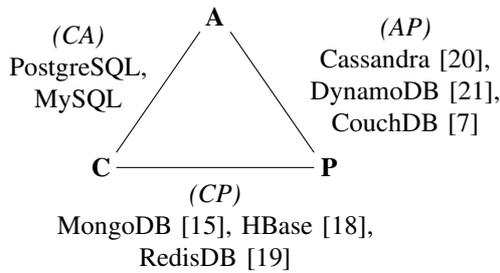
\begin{figure}[t]
    \centering
    \begin{tikzpicture}
    \node at (0,0) (C){\textbf{C}};
    \node at (1.5,2) (A){\textbf{A}};
    \node at (3,0) (P){\textbf{P}};
    
    \draw (0.2,0) -- (2.8,0); 
        \node at (1.5, -0.8) 
        {\makecell{\textit{(CP)} \\ MongoDB~\cite{MongoDB17}, HBase~\cite{Vora2011}, \\ RedisDB~\cite{carlson2013}}};
    
    \draw (0.2,0.2) -- (1.3,1.8);    
        \node at (4,1.2)
        {\makecell{\textit{(AP)} \\ Cassandra~\cite{Lakshman2010}, \\ DynamoDB~\cite{Sivasubramanian2012}, \\ CouchDB~\cite{Anderson2010}}};
    
    \draw (1.7,1.8) -- (2.8,0.2);
        \node at (-0.3,1.3)
        {\makecell{\textit{(CA)} \\ PostgreSQL, \\ MySQL}};
    \end{tikzpicture}
    \vspace{-0.4cm}
    \caption{CAP Triangle for Database systems}
    \label{fig:CAP}
    \vspace{-0.4cm}
\end{figure}

CAP theorem has also made its influence in the blockchain realm (see, for example \cite{weber2017availability}). If we pick Availability over Consistency, any reads are not guaranteed to be up-to-date, and we call the system as AP. However, if we choose Consistency over Availability, the system, called CP, would be unavailable at the time of partition and might disrupt the consensus. Thus in blockchain systems, both properties are desirable. Though blockchain does not always require strong consistency, eventual consistency can serve the purpose and can be achieved through consensus. For example, in the case of bitcoin, the longest chain method brings eventual consistency, but there are no fix methods to achieve eventual consistency and leaves this topic for debate. Figure~\ref{fig:CAP} shows the different database systems according to the CAP theorem. 

An analogy to the CAP theorem for blockchain have been proposed as the DCS theorem~\cite{zhang2018}, where DCS abbreviation refers to \textbf{D}ecentralization, \textbf{C}onsistency, \textbf{S}calability. The DCS theorem states that a blockchain system can have at most two properties simultaneously out of the three estates of DCS. The DCS properties can be defined as follows:
\begin{itemize}
    \item \textit{Decentralization} - There is no trusted entity controlling the network, hence no single point of failure. Blockchains are inherently decentralized, but in the DCS triangle, we are considering the case of full decentralization. In the case of full decentralization, any node can join the network and participate as a validator.
    \item \textit{Consistency} - The blockchain nodes will read the same data at the same time. The query for the blockchain data on any blockchain node should fetch the same result. The consistency in blockchain should prevent double-spending and should be brought from the consensus algorithm used. 
    \item \textit{Scalability} - The performance of blockchain should increase with the increase in the number of peers and the number of allocated computational resources. The throughput and the capacity of the system should be high, and latency should be low.
\end{itemize}

\begin{figure}[t]
    \centering
    \begin{tikzpicture}
    \node at (0,0) (C){\textbf{C}};
    \node at (1.5,2) (S){\textbf{S}};
    \node at (3,0) (D){\textbf{D}};
    
    \draw (0.2,0) -- (2.8,0); 
        \node at (1.5, -0.6) 
        {\makecell{\textit{(DC)} \\ Bitcoin~\cite{Nakamoto_bitcoin}, Ethereum~\cite{Ethereum}}};
    
    \draw (0.2,0.2) -- (1.3,1.8);    
        \node at (3.2,1.4)
        {\makecell{\textit{(DS)} \\ IPFS~\cite{IPFS}}};
    
    \draw (1.7,1.8) -- (2.8,0.2);
        \node at (-0.8,1.3)
        {\makecell{\textit{(CS)} \\ Hyperledger~\cite{Androulaki2018}, \\ MultiChain~\cite{MultiChain_web}}};
    \end{tikzpicture}  
    \vspace{-0.4cm}
    \caption{DCS Triangle for Blockchain systems}
    \label{fig:DCS}
    \vspace{-0.5cm}
\end{figure}
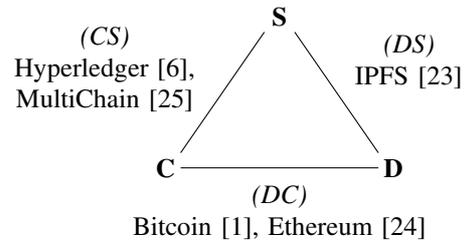

In a similar way to CAP, we can also categorize the blockchain systems in DCS as DC, CS, and DS systems as trade-offs between the DCS properties. Most of the crypto-currencies like Bitcoin~\cite{Nakamoto_bitcoin} can be considered as DC systems. Nevertheless, all the permissioned blockchains do not have full decentralization, hence should be regarded as CS systems. Systems like Interplanetary File System (IPFS)~\cite{IPFS} do not provide consistency as the different parts of data are distributed to different nodes (thus, they are DS systems). Figure~\ref{fig:DCS} depicts the different systems, according to the DCS theorem.

If we apply a similar relaxation approach as it was used for the proof of the CAP theorem in \cite{gilbert2002brewer}, we have the following reasoning: In DC systems, scalability is a big issue. Hence, to solve the scalability, many techniques are proposed, such as Sharding~\cite{Sharding}, Lightening network~\cite{poon2016bitcoin}, or by using the scalable consensus algorithms. Furthermore, in DS systems, the consistency can be achieved by using the safe and verifiable smart contracts, by making the blockchain attack resilient and by handling the forks. Therefore in a way, all the DCS properties are achievable with some appropriate relaxations and balances. Here for blockchain systems, we postulate the following conjecture for achieving all three properties:
\begin{conjecture}[DCS-satisfiability]
There exist a well-balanced and relaxed set of requirements for Decentralization, Consistency, and Scalability (DCS) properties such that a blockchain system can have all three properties satisfied.
\end{conjecture}

While for the CAP theorem, the relaxation of the requirements was achieved by the introduction of the $t$-connected consistency model,  a precise analogous mathematical modeling for the blockchain systems is an active and open field of research.

\begin{table*}[th]
    \resizebox{\textwidth}{!}{
    \centering
    \begin{tabular}{|l|c|c|c|c|c|c|c|c|c|}
    \hline
    \textbf{System} & \textbf{Data Model} & \textbf{Consensus} & \textbf{\makecell{Decentral-\\ ization}} & \textbf{Consistency} & \textbf{Scalability} & \textbf{Immutability} & \textbf{\makecell{Low \\ Latency}} & \textbf{\makecell{High \\ Throughput}} & \textbf{Sharding} \\
    \hline
    \hline
    
    \textbf{BigchainDB} & \makecell{MongoDB, \\ RethinkDB} & Tendermint\cite{kwon2014tendermint} & \cmark & \cmark & \cmark & \cmark & \cmark & \cmark & \cmark \\
    \hline
    
    \textbf{BlockchainDB} & Key-Value & \makecell{Underlying \\ Blockchain \\ Consensus} & \cmark & \cmark & --- & \cmark & \xmark & \xmark & \cmark \\
    \hline
    
    \textbf{Cassandra} & \makecell{Key-Value, \\ Column store} & \makecell{Paxos~\cite{lamport1998} \\ Consensus} & \cmark & \cmark* & \cmark & \xmark & \cmark & \cmark & \cmark-- \\
    \hline
    
    \textbf{ChainifyDB} & \makecell{Relational} & \makecell{Whatever \\ ledger \\ Consensus} & \cmark & --- & \cmark & \cmark & --- & \cmark & --- \\
    \hline
    
    \textbf{CockroachDB} & Key-Value & \makecell{Raft \\ Consensus} & \xmark & \cmark & \cmark & \xmark & \cmark & --- & \cmark \\
    \hline
    
    \textbf{CosmosDB} & \makecell{Key-Value, \\ Document, \\ Graph} & \makecell{No \\ Consensus} & \xmark & \cmark** & \cmark & \xmark & \cmark & \cmark & \cmark \\
    \hline
    
    \textbf{CouchDB} & Key-Value & \makecell{No \\ Consensus} & \xmark & \cmark* & \cmark & \xmark & \cmark & \cmark & \cmark-- \\
    \hline
    
    \textbf{CovenantSQL} & \makecell{SQLite DB} & \makecell{DPOS~$^1$, \\ BFT-Raft~\cite{clow2017}} & \cmark & \cmark & --- & \cmark & --- & --- & --- \\
    \hline
    
    \textbf{Dqlite} & \makecell{SQLite DB} &  \makecell{C-Raft~\cite{Diego2014} \\ Consensus} & \xmark & --- & --- & \xmark & \cmark & \cmark & --- \\
    \hline
    
    \textbf{FlureeDB} & \makecell{Document, \\ Graph} & \makecell{PBFT, \\ Raft} &  \cmark & \cmark & \cmark & \cmark & \cmark & \cmark & \cmark \\
    \hline 
    
    \textbf{HBasechainDB} & HBase & \makecell{No, but uses\\ Blockchain \\ Pipelining} &  \cmark & \cmark & \cmark & \cmark & \cmark & \cmark & \cmark \\
    \hline 
    
    \textbf{MongoDB} & \makecell{Document \\ Based} & \makecell{Raft \\ Based} & \xmark & \cmark & \cmark & \xmark & --- & \xmark & \cmark \\
    \hline 
    
    \textbf{OurSQL} & Mysql & \makecell{POW type \\ Consensus} & \cmark & --- & --- & \cmark & --- & --- & ---\\
    \hline
    
    \textbf{Postchain} & \makecell{Relational} & \makecell{BFT \\ Based} & \cmark & \cmark & \cmark & \cmark  & --- & --- & ---\\ 
    \hline
    
    \textbf{ProvenDB} & \makecell{MongoDB} & \makecell{Not \\ Mentioned} & \cmark & \cmark & --- & \cmark & --- & \xmark & ---\\
    \hline
    
    \textbf{QLDB} & \makecell{Document} & \makecell{No \\ Consensus} & \xmark & \cmark & \cmark & \cmark & --- & --- & --- \\
    \hline
    
    \textbf{Rqlite} & \makecell{SQLite DB} &  \makecell{Raft \\ Consensus} & \xmark & \cmark** & --- & \xmark & --- & --- & --- \\
    \hline
    
    \textbf{TiesDB} & \makecell{Document} & \makecell{BFT \\ Based} & \cmark & --- & --- & \xmark & --- & --- & \cmark \\
    \hline
    
    \textbf{TitaniumDB} & Key-Value & \makecell{Raft \\ Consensus} & \xmark & \cmark & \cmark & \xmark & --- & --- & \cmark \\
    \hline
    
    \textbf{VoltDB} & \makecell{Relational} & \makecell{No \\ Consensus} & \xmark & \cmark & \cmark & \xmark & \cmark & \cmark & \cmark \\
    \hline
    
    \end{tabular}
    }
    \caption{Comparison Matrix for Different Systems. Here, `\cmark' indicates that the feature is present, `\xmark' indicates that the feature is not present in the corresponding system, `\cmark*' represents eventual consistency, `\cmark**' represents configurable consistency, `\cmark--' represents that database has its own sharding method, `---' represents inconclusive data}
    \label{tab:comparison}
    \vspace{-0.75cm}
\end{table*}

\section{Databases for blockchain use} \label{Traditional-DB}
The database systems have been used for storing transaction data of blockchain. The following databases have different characteristics. Based on these characteristics, a database can be chosen to be used in particular blockchain applications.

\subsection{Relational Database Systems}

\textbf{PostgreSQL}~\cite{Postgresql} is a free and open-source relational database management system (RDBMS). It has a wide variety of native data types and supports user-defined objects, which can be beneficial to define blockchain assets in the blockchain system. It is highly modular, extensible, and also supports isolation on different levels. PostgreSQL has been used to create \textit{blockchain relational database}, where the replicas are managed by different organizations that do not trust each other~\cite{Nathan2019}. 

\textbf{MySQL}~\cite{MariaDB} and its community developed fork \textbf{MariaDB} is open source relational database system with advanced replication and clustering features. \textbf{OurSQL}~\footnote{\url{https://en.bitcoinwiki.org/wiki/DPoS},  \url{http://oursql.org}, \url{https://covenantsql.io}, \url{https://dqlite.io}, \url{http://www.rqlite.com/}} is a standalone server connected to MySQL database. It is a combination of Blockchain and MySQL. OurSQL can be used for private blockchain applications.

\textbf{SQLite}~\cite{SQLite} is an embedded, non client-server, ACID-compliant relational database system. It is suitable to be embedded as a local database in the blockchain nodes.

\textbf{CovenantSQL} (CQL)~$^1$ is a decentralized, trusted, GDPR-compliant with blockchain features built on SQLite.
    It can be used as a low cost database as a service (DBaaS). CQL has layered architecture, consisting of Global Consensus Layer, SQL Consensus Layer, and Datastore Layer. 
\textbf{Dqlite}~$^1$ (“distributed SQLite”) is an open-source, fast, Disk-backed database with in-memory options. 
It best suits for fault-tolerant IoT and Edge devices. \textbf{RQLITE}~$^1$ is an open-source, lightweight, fault-tolerant, and distributed relational database.
It allows the dynamic creation of a cluster of nodes and provides node-to-node encryption. RQLITE appears a potential candidate for the lightweight blockchain solutions. 

\subsection{NoSQL Database Systems}
\textbf{MongoDB}~\cite{MongoDB17} 
is the fastest-growing document-based database in the market. The distributed architecture of MongoDB makes it an ideal platform for building blockchain databases. MongoDB offers data model flexibility, high scalability, robust security, complex queries, and SQL capabilities. Due to its powering technological features, it is used by many leading enterprises nowadays. The MongoDB Enterprise edition supports encryption, auditing, sharding, and access control. \textbf{BigchainDB}~\cite{BigchainDB} was initially built upon RethinkDB~\cite{walsh2009} cluster, but from version 2.0 it employs Tendermint consensus~\cite{kwon2014tendermint} over a set of independent MongoDB instances. Also, \textbf{ProvenDB}~\cite{ProvenDB} adds the blockchain characteristic layer on top of the MongoDB database. \textbf{EthernityDB}~\cite{Helmer2018} can be used to integrate database functionalities in Ethereum blockchain~\cite{Ethereum} by modularizing the Ethereum smart contracts. EthernityDB uses MongoDB for the coupling with Ethereum blockchain.

\textbf{CouchDB}~\cite{Anderson2010} is a key-value data store and provides rich query capability similar to MongoDB. Hyperledger Fabric~\cite{Androulaki2018} uses CouchDB as a state database for storing chaincode processed transaction data as key-value pairs. It supports rich queries against chaincode data. 
Hyperledger Composer also uses CouchDB by converting SQL queries into CouchDB JSON queries.

\textbf{QLDB} Amazon QLDB~\cite{QLDB} is a ledger database that abstracts many features of blockchain. It renders a tamper-resistant, transparent, and cryptographically verifiable ledger of transactions. QLDB lacks decentralization and hence does not follow any consensus algorithm. Therefore it best suits the enterprises which do not require any consensus and still want to have immutability of its data. QLDB also supports SQL queries.

\textbf{Cassandra} Apache Cassandra~\cite{Lakshman2010} is one of the most popular NoSQL database developed by Facebook. Currently, it is in use at many big enterprises like Netflix, Instagram, Github, and eBay. It is a fully decentralized system and provides great performance, durability and fault tolerance without compromising availability. Cassandra has its query language named CQL to interact with the system. When establishing consistency, Cassandra also supports lightweight transactions.

\textbf{TiesDB}~\cite{TiesDB17} is a public, decentralized, and distributed database. Ties Network is a deep modification of the Cassandra database. It is flexible on choosing an underlying NoSQL database and hence inherits most of the features from the underlying database. It adds Byzantine fault tolerance (BFT), while most of the NoSQL database lacks BFT. It is fast and supports sharding, smart contracts, and incentive schemes. It can be used to build decentralized applications providing fast data retrieval. 

\textbf{CosmosDB} Azure Cosmos~\cite{GuayPaz2018} DB is Microsoft's globally distributed, fully-decentralized, multi-model database. The database models can be key-value, graph, or document. 
It provides availability and consistency with comprehensive service level agreements (SLAs). It offers multi-master replication across various regional distributions. Many enterprises can benefit from building a decentralized blockchain application using CosmosDB.

\textbf{HBase}~\cite{Hbase} is a NoSQL distributed database tuned for the massive data sets. \textbf{HBasechainDB}~\cite{Sahoo2018} is a scalable big data store based on the concept of blockchain. This framework gives the ability to handle the big data of blockchain. HBaseChainDb works on the underlying HBase database~\cite{Vora2011} in the Hadoop ecosystem. It adds blockchain functionalities of decentralization and immutability on the top of HBase. HBaseChainDB can be used by enterprises whose systems already exist in the Hadoop ecosystem.

Some of the promising NoSQL databases for the use in blockchain are RethinkDB~\cite{walsh2009}, RedisDB~\cite{carlson2013}, AWS DynamoDB~\cite{Sivasubramanian2012}, and Etcd~\cite{Etcd}.

\subsection{NewSQL Database Systems} Those are relational, distributed database systems that offer ACID semantics without compromising the scalability. After the introduction of Google Spanner~\cite{Spanner}, the first NewSQL system, many NewSQL systems evolved. These NewSQL systems can be useful for building blockchain applications. Some of the promising NewSQL systems are as follows:

    \textbf{VoltDB}~\cite{Stonebraker2013} is an open-source, in-memory database. The new version of VoltDB V8 adds many new capabilities. It supports user-defined SQL functions, which can be useful in smart contracts in blockchain. It provides SQL support for the traversal of blockchain records with recursive Common Table Expressions (CTE). 
    
    \textbf{TiDB}~\cite{TiDB} (``Titanium DB") is an open-source, distributed SQL database with strong consistency and high availability. It has modular design containing three components for the cluster coordination, replicating key-value store, and scheduling SQL queries. 
    
    \textbf{CockroachDB}~\cite{CockroachDB} is an open-source, key-value database. It supports strongly consistent ACID semantic and horizontal scalability. It uses 2-phase commit protocols for transaction serializability. 

\subsection{Modern databases influenced by blockchain} \label{others}

\textbf{FlureeDB}~\cite{FlureeDB} is a scalable blockchain database. It consolidates blockchain with the document and graph database to support a broad range of industrial use cases. It provides rich access capability directly inside the database. It offers sharding, censorship resistance, privacy, cloud hosting, and uses composite consensus, which enables multiple DBs to be queried as a single DB. Each block of blockchain in FlureeDB represents a unique time moment, and this feature is called  "time-travel." 
Many of the enterprise applications with complex data-needs can benefit from FlureeDB and its features.

\textbf{Postchain}~\cite{Postchain} combines the features of a mature distributed database and blockchain. A blockchain solution can be implemented using Postchain and SQL. 
It has powerful features to manage integrity, validation, and data independence, along with the inherited traits from the underlying standard database.

\textbf{BlockchainDB}~\cite{El-Hindi2019} implements a database layer on the top of an existing blockchain system and leverages database system capabilities like SQL queries. It provides partitioned storage of blockchain data among the peers in the network.

\textbf{ChainifyDB}~\cite{schuhknecht2019chainifydb} adds the blockchain characteristic layer on top of a standard database. Hence, it leverages enterprises to build decentralized cutting-edge blockchain applications on top of their database systems. 

There are many other solutions for providing different functionalities in blockchain systems. For example, \textbf{OrbitDB}~\cite{OrbitDB} can be an excellent choice for blockchain applications or decentralized apps (dApps). Moreover, some of the works are also oriented towards providing a specific functionality in the blockchain. For example,  \textbf{JainDB}~\footnote{\url{https://github.com/rzander/jaindb}}, is a data warehouse for JSON objects and provides REST API services to interact with the blockchain data store;  \textbf{vChain}~\cite{vchain}, \textbf{VQL}~\cite{VQL}, delivers efficient and verifiable data query services in the blockchain systems. The analysis presented in this section is summarized in Table \ref{tab:comparison}. 

\section{conclusion} \label{Conclusion}
\vspace{-0.2cm}
The last decade was a decade of an intense interchanged and mutually influenced development of the database and blockchain technologies.  To the best of our knowledge, this is the first systematized study of those development trends. We provided a detailed summary of traditional databases, which are used or can be used in the design of blockchain platforms or applications. Further, we presented a detailed explanation of different decentralized solutions that uses traditional databases and provides blockchain-enabled solutions. We also discussed the analogous theorem to the CAP theorem for the databases known as DCS theorem and postulated an analogous DCS-satisfiability conjecture.
\vspace{-0.2cm}
\bibliographystyle{IEEEtran}
\bibliography{report}

\begin{thebibliography}{10}
\providecommand{\url}[1]{#1}
\csname url@samestyle\endcsname
\providecommand{\newblock}{\relax}
\providecommand{\bibinfo}[2]{#2}
\providecommand{\BIBentrySTDinterwordspacing}{\spaceskip=0pt\relax}
\providecommand{\BIBentryALTinterwordstretchfactor}{4}
\providecommand{\BIBentryALTinterwordspacing}{\spaceskip=\fontdimen2\font plus
\BIBentryALTinterwordstretchfactor\fontdimen3\font minus
  \fontdimen4\font\relax}
\providecommand{\BIBforeignlanguage}[2]{{%
\expandafter\ifx\csname l@#1\endcsname\relax
\typeout{** WARNING: IEEEtran.bst: No hyphenation pattern has been}%
\typeout{** loaded for the language `#1'. Using the pattern for}%
\typeout{** the default language instead.}%
\else
\language=\csname l@#1\endcsname
\fi
#2}}
\providecommand{\BIBdecl}{\relax}
\BIBdecl

\bibitem{Nakamoto_bitcoin}
S.~Nakamoto, ``Bitcoin: A peer-to-peer electronic cash system,
  http://bitcoin.org/bitcoin.pdf,'' 2009.

\bibitem{Chowdhury2018}
M.~J.~M. {Chowdhury}, A.~{Colman}, M.~A. {Kabir}, J.~{Han}, and P.~{Sarda},
  ``Blockchain versus database: A critical analysis,'' in \emph{2018 IEEE
  TrustCom/BigDataSE}, Aug 2018, pp. 1348--1353.

\bibitem{brewer2000towards}
E.~A. Brewer, ``Towards robust distributed systems,'' in \emph{PODC},
  vol.~7.\hskip 1em plus 0.5em minus 0.4em\relax Portland, OR, 2000.

\bibitem{FlureeDB}
\BIBentryALTinterwordspacing
``Flureedb,'' 2017. [Online]. Available: \url{https://flur.ee}
\BIBentrySTDinterwordspacing

\bibitem{BigchainDB}
T.~McConaghy \emph{et~al.}, ``{BigchainDB: a scalable blockchain database},''
  \emph{{White paper}}, 2016.

\bibitem{Androulaki2018}
E.~Androulaki \emph{et~al.}, ``Hyperledger fabric: A distributed operating
  system for permissioned blockchains,'' in \emph{Proceedings of the Thirteenth
  EuroSys Conference}.\hskip 1em plus 0.5em minus 0.4em\relax ACM, 2018, pp.
  30:1--30:15.

\bibitem{Anderson2010}
J.~C. Anderson, J.~Lehnardt, and N.~Slater, \emph{CouchDB: the definitive
  guide: time to relax}.\hskip 1em plus 0.5em minus 0.4em\relax O'Reilly Media,
  Inc., 2010.

\bibitem{VQL}
Z.~{Peng}, H.~{Wu}, B.~{Xiao}, and S.~{Guo}, ``{VQL: Providing Query Efficiency
  and Data Authenticity in Blockchain Systems},'' in \emph{2019 IEEE ICDEW},
  April 2019, pp. 1--6.

\bibitem{Bergman2018}
S.~Bergman, M.~Asplund, and S.~Nadjm-Tehrani, ``Permissioned blockchains and
  distributed databases: A performance study,'' \emph{Concurrency and
  Computation: Practice and Experience}, p. e5227, 2018.

\bibitem{raikwar2019sok}
M.~Raikwar, D.~Gligoroski, and K.~Kralevska, ``{SoK of used cryptography in
  blockchain},'' \emph{IEEE Access}, vol.~7, pp. 148\,550--148\,575, 2019.

\bibitem{hasselgren2019blockchain}
A.~Hasselgren, K.~Kralevska, D.~Gligoroski, S.~A. Pedersen, and A.~Faxvaag,
  ``Blockchain in healthcare and health sciences--a scoping review,''
  \emph{International Journal of Medical Informatics}, p. 104040, 2019.

\bibitem{gilbert2002brewer}
S.~Gilbert and N.~Lynch, ``Brewer's conjecture and the feasibility of
  consistent, available, partition-tolerant web services,'' \emph{Acm Sigact
  News}, vol.~33, no.~2, pp. 51--59, 2002.

\bibitem{Kung1981}
H.~T. Kung and J.~T. Robinson, ``On optimistic methods for concurrency
  control,'' \emph{ACM Trans. Data. Syst.}, vol.~6, no.~2, pp. 213--226, Jun.
  1981.

\bibitem{El-Hindi2019}
M.~El-Hindi, M.~Heyden, C.~Binnig, R.~Ramamurthy, A.~Arasu, and D.~Kossmann,
  ``{BlockchainDB - Towards a Shared Database on Blockchains},'' in
  \emph{Proceedings of the 2019 International Conference on Management of
  Data}.\hskip 1em plus 0.5em minus 0.4em\relax ACM, 2019, pp. 1905--1908.

\bibitem{MongoDB17}
MangoDB, ``Building enterprise-grade blockchain databases with mongodb,''
  \emph{White paper}, Nov 2017.

\bibitem{Bernstein1983}
P.~A. Bernstein and N.~Goodman, ``Multiversion concurrency control—theory and
  algorithms,'' \emph{ACM Trans. Data. Syst.}, vol.~8, no.~4, pp. 465--483,
  1983.

\bibitem{weber2017availability}
I.~Weber, V.~Gramoli, A.~Ponomarev, M.~Staples, R.~Holz, A.~B. Tran, and
  P.~Rimba, ``On availability for blockchain-based systems,'' in \emph{2017
  IEEE SRDS}, 2017, pp. 64--73.

\bibitem{Vora2011}
{Mehul Nalin Vora}, ``Hadoop-hbase for large-scale data,'' in \emph{Proceedings
  of 2011 International Conference on Computer Science and Network Technology},
  vol.~1, Dec 2011, pp. 601--605.

\bibitem{carlson2013}
J.~L. Carlson, \emph{Redis in action}.\hskip 1em plus 0.5em minus 0.4em\relax
  Manning Shelter Island, 2013.

\bibitem{Lakshman2010}
A.~Lakshman and P.~Malik, ``Cassandra: A decentralized structured storage
  system,'' \emph{SIGOPS Oper. Syst. Rev.}, vol.~44, no.~2, pp. 35--40, Apr.
  2010.

\bibitem{Sivasubramanian2012}
S.~Sivasubramanian, ``{Amazon dynamoDB: a seamlessly scalable non-relational
  database service},'' in \emph{Proceedings of the 2012 ACM SIGMOD
  International Conference on Management of Data}, 2012, pp. 729--730.

\bibitem{zhang2018}
K.~Zhang and H.-A. Jacobsen, ``Towards dependable, scalable, and pervasive
  distributed ledgers with blockchains.'' in \emph{ICDCS}, 2018, pp.
  1337--1346.

\bibitem{IPFS}
\BIBentryALTinterwordspacing
J.~Benet, ``{IPFS} - content addressed, versioned, {P2P} file system,''
  \emph{arXiv}, 2014. [Online]. Available: \url{http://arxiv.org/abs/1407.3561}
\BIBentrySTDinterwordspacing

\bibitem{Ethereum}
G.~Wood, ``{Ethereum: A Secure Decentralised Generalised Transaction Ledger},''
  Yellow Paper, 2014.

\bibitem{MultiChain_web}
\BIBentryALTinterwordspacing
``Multichain,'' 2015. [Online]. Available: \url{https://www.multichain.com/}
\BIBentrySTDinterwordspacing

\bibitem{Sharding}
L.~Luu, V.~Narayanan, C.~Zheng, K.~Baweja, S.~Gilbert, and P.~Saxena, ``{A
  Secure Sharding Protocol For Open Blockchains},'' in \emph{Proceedings of the
  2016 ACM SIGSAC Conference on Computer and Communications Security}.\hskip
  1em plus 0.5em minus 0.4em\relax ACM, 2016, pp. 17--30.

\bibitem{poon2016bitcoin}
J.~Poon and T.~Dryja, ``{The Bitcoin Lightning Network: Scalable off-chain
  instant payments},''
  \url{https://www.bitcoinlightning.com/wp-content/uploads/2018/03/lightning-network-paper.pdf},
  2016.

\bibitem{kwon2014tendermint}
\BIBentryALTinterwordspacing
J.~Kwon, ``Tendermint: Consensus without mining,'' \emph{Draft v. 0.6, fall},
  vol.~1, p.~11, 2014. [Online]. Available: \url{https://tendermint.com}
\BIBentrySTDinterwordspacing

\bibitem{lamport1998}
L.~Lamport, ``The part-time parliament,'' \emph{ACM Transactions on Computer
  Systems (TOCS)}, vol.~16, no.~2, pp. 133--169, 1998.

\bibitem{clow2017}
\BIBentryALTinterwordspacing
J.~Clow and Z.~Jiang, ``A byzantine fault tolerant raft,'' 2017. [Online].
  Available:
  \url{https://www.scs.stanford.edu/17au-cs244b/labs/projects/clow_jiang.pdf}
\BIBentrySTDinterwordspacing

\bibitem{Diego2014}
D.~Ongaro and J.~Ousterhout, ``In search of an understandable consensus
  algorithm,'' in \emph{2014 {USENIX} Annual Technical Conference}.\hskip 1em
  plus 0.5em minus 0.4em\relax Philadelphia, PA: {USENIX} Association, Jun.
  2014, pp. 305--319.

\bibitem{Postgresql}
\BIBentryALTinterwordspacing
``Postgresql v10.'' [Online]. Available: \url{https://www.postgresql.org}
\BIBentrySTDinterwordspacing

\bibitem{Nathan2019}
S.~Nathan, C.~Govindarajan, A.~Saraf, M.~Sethi, and P.~Jayachandran,
  ``Blockchain meets database: Design and implementation of a blockchain
  relational database,'' \emph{arXiv}, 2019.

\bibitem{MariaDB}
\BIBentryALTinterwordspacing
{MariaDB Foundation }, ``{MariaDB 10.5.0 now available},'' Dec 2019. [Online].
  Available: \url{https://mariadb.org/mariadb-10-5-0-now-available/}
\BIBentrySTDinterwordspacing

\bibitem{SQLite}
\BIBentryALTinterwordspacing
{Hipp, Wyrick \& Company, Inc.}, ``{SQLite},'' 2020. [Online]. Available:
  \url{https://www.sqlite.org/}
\BIBentrySTDinterwordspacing

\bibitem{walsh2009}
\BIBentryALTinterwordspacing
L.~Walsh, V.~Akhmechet, and M.~Glukhovsky, ``Rethinkdb-rethinking database
  storage,'' 2009. [Online]. Available:
  \url{https://pdfs.semanticscholar.org/3cdf/b12ceee0f82a08b352cead0bf791477dca98.pdf}
\BIBentrySTDinterwordspacing

\bibitem{ProvenDB}
{ProvenDB Team}, ``{ProvenDB, Trust your data},'' \emph{Yellow paper}, 2019.

\bibitem{Helmer2018}
S.~Helmer, M.~Roggia, N.~E. Ioini, and C.~Pahl, ``{EthernityDB -- Integrating
  Database Functionality into a Blockchain},'' in \emph{New Trends in Databases
  and Information Systems}.\hskip 1em plus 0.5em minus 0.4em\relax Springer
  International Publishing, 2018, pp. 37--44.

\bibitem{QLDB}
\BIBentryALTinterwordspacing
``Amazon quantum ledger database,'' 2019. [Online]. Available:
  \url{https://aws.amazon.com/qldb/}
\BIBentrySTDinterwordspacing

\bibitem{TiesDB17}
D.~K. Anton~Filatov, ``Ties.network, database management system, technical
  description,'' \emph{Yellow paper}, Aug 2017.

\bibitem{GuayPaz2018}
J.~R. Guay~Paz, \emph{Introduction to Azure Cosmos DB}.\hskip 1em plus 0.5em
  minus 0.4em\relax Berkeley, CA: Apress, 2018, pp. 1--23.

\bibitem{Hbase}
\BIBentryALTinterwordspacing
{The Apache Software Foundation}, ``{Welcome to Apache HBase™},'' 2020.
  [Online]. Available: \url{https://hbase.apache.org/}
\BIBentrySTDinterwordspacing

\bibitem{Sahoo2018}
M.~S. Sahoo and P.~K. Baruah, ``Hbasechaindb -- a scalable blockchain framework
  on hadoop ecosystem,'' in \emph{Supercomputing Frontiers}, R.~Yokota and
  W.~Wu, Eds.\hskip 1em plus 0.5em minus 0.4em\relax Cham: Springer
  International Publishing, 2018, pp. 18--29.

\bibitem{Etcd}
\BIBentryALTinterwordspacing
``Etcd: A distributed, reliable key-value store for the most critical data of a
  distributed system,'' 2019. [Online]. Available: \url{https://etcd.io}
\BIBentrySTDinterwordspacing

\bibitem{Spanner}
J.~C. Corbett, J.~Dean, M.~Epstein, A.~Fikes, C.~Frost, J.~J. Furman,
  S.~Ghemawat, A.~Gubarev, C.~Heiser, P.~Hochschild \emph{et~al.}, ``Spanner:
  Google’s globally distributed database,'' \emph{ACM Transactions on
  Computer Systems (TOCS)}, vol.~31, no.~3, pp. 1--22, 2013.

\bibitem{Stonebraker2013}
M.~Stonebraker and A.~Weisberg, ``{The VoltDB Main Memory DBMS},'' \emph{IEEE
  Data Eng. Bull.}, vol.~36, no.~2, pp. 21--27, 2013.

\bibitem{TiDB}
``{TiDB, SQL at Scale},'' \url{https://pingcap.com/en/}, 2019.

\bibitem{CockroachDB}
\BIBentryALTinterwordspacing
``{CockroachDB, An evolution of the database},'' 2017. [Online]. Available:
  \url{https://www.cockroachlabs.com}
\BIBentrySTDinterwordspacing

\bibitem{Postchain}
J.~M. Graglia and C.~Mellon, ``Blockchain and property in 2018: At the end of
  the beginning,'' \emph{Innovations: Technology, Governance, Globalization},
  vol.~12, no. 1-2, pp. 90--116, 2018.

\bibitem{schuhknecht2019chainifydb}
F.~M. Schuhknecht, A.~Sharma, J.~Dittrich, and D.~Agrawal, ``{ChainifyDB: How
  to Blockchainify any Data Management System},'' 2019.

\bibitem{OrbitDB}
\BIBentryALTinterwordspacing
``{OrbitDB, Peer-to-Peer Databases for the Decentralized Web},'' 2015.
  [Online]. Available: \url{https://orbitdb.org}
\BIBentrySTDinterwordspacing

\bibitem{vchain}
C.~Xu, C.~Zhang, and J.~Xu, ``{vChain: Enabling Verifiable Boolean Range
  Queries over Blockchain Databases},'' in \emph{Proceedings of the 2019
  International Conference on Management of Data}, 2019, pp. 141--158.

\end{thebibliography}

\end{document}